\documentclass[aps,twocolumn,amsmath,amssymb,floatfix,showpacs,nofootinbib]{revtex4}
\usepackage{graphicx} \usepackage{epsfig}

\usepackage{amsbsy,bm}

\renewcommand{\H}{H$_2$~}
  
  \newcommand{\Hp}{H$_2^+$~}
	\newcommand{\water}{H$_2$O~}
\newcommand{\bxi}{\boldsymbol{\xi}}
\newcommand{\vecr}{\mathbf{r}}
\newcommand{\ba}{\begin{eqnarray}}
\newcommand{\ea}{\end{eqnarray}}
\newcommand{\br}{\begin{eqnarray*}}
\newcommand{\er}{\end{eqnarray*}}
\newcommand{\be}{\begin{equation}}
\newcommand{\ee}{\end{equation}}

\newcommand{\eref}[1] {(\ref{#1})}
\newcommand{\Eref}[1] {Eq.~(\ref{#1})}
\newcommand{\Fref}[1] {Fig. \ref{#1}}
\newcommand{\Tref}[1] {Table \ref{#1}}

\newcommand{\np}{\newpage}

\begin{document}

\title{Time delay in XUV/IR photoionization of \water}

\author{Vladislav V. Serov}
\affiliation{
Department of Theoretical Physics, Saratov State University, 83
Astrakhanskaya, Saratov 410012, Russia}

\author{A. S. Kheifets}

\affiliation{Research School of Physics and Engineering,
The Australian National University,
Canberra ACT 2601, Australia}

\date{\today}
\begin{abstract}
We solve the time-dependent Schr\"odinger equation describing a
water molecule driven by a superposition of the XUV and IR pulses
typical for a RABBITT experiment. This solution is obtained by a
combination of  the time-dependent coordinate scaling  and the density
functional theory with self-interaction correction. Results of this
solution are used to determine the time delay in photoionization of the
water and hydrogen molecules.
\end{abstract}

\pacs{33.20.Xx, 33.80.Eh, 33.80.-b}


\maketitle

\section{Introduction}

Time delay in molecular photoionization can now be measured by using
the RABBITT technique (Reconstruction of Attosecond Beating By
Interference of Two-photon Transitions). In the pioneering experiment,
\citet{PhysRevLett.117.093001} determined the relative delay in
photoemission of the outer-most valence shells of the H$_2$O and
N$_2$O molecules. The Eisenbud-Wigner-Smith component of the measured
time delay (Wigner time delay for brevity), related to the XUV photon
absorption, was evaluated from the complex dipole matrix elements
provided by molecular quantum scattering theory
\cite{Baykusheva2017}. The Coulomb-laser coupling (CLC) correction,
associated with the IR dressing field, was accounted for separately in
an atomic-like fashion.

In the present work, we provide an alternative approach where the
ionizing XUV field and the dressing IR field drive the molecular
time-dependent Schr\"odinger equation (TDSE). This equation is solved
by a combination of the time-dependent coordinate scaling (TDCS) and
the density functional theory with self-interaction correction
(DFT-SIC). We applied the TDCS technique in our earlier work to
describe a RABBITT measurement on the molecular \Hp ion
\cite{PhysRevA.93.063417}. An advantage of this technique is that the
TDSE is solved directly in the XUV and IR fields and the calculated
(and measurable) time delay is not artificially split into
the Wigner and CLC components. The earlier application to \Hp was
based on the explicit form of the one-electron potential. For more
complex molecules like \H and \water, this potential can be  evaluated
within the DFT-SIC approach.


\section{Theoretical methods}

\subsection{Time-dependent scaling method}

We restrict ourselves with a single active electron (SAE)
approximation and write the TDSE as
\begin{eqnarray}
i\frac{\partial\Psi(\vecr,t)}{\partial
t}=\hat{H}\Psi(\vecr,t)  
\label{TDSE}
\end{eqnarray}
with the Hamiltonian
\begin{eqnarray}
\hat{H}=\frac{\hat{p}^{\,2}}{2}-\mathbf{A}(t)\hat{\mathbf{p}}+U(\vecr).
\end{eqnarray}
Here $\hat{\mathbf{p}}=-i\nabla$ is the momentum operator,
$U(\vecr,t)$ is the electron-nucleus interaction, $\mathbf{A}(t)$
is the vector potential of the electromagnetic field. The latter is
defined as
\footnote{The atomic units are in use throughout the paper such that $e=m=\hbar=1$.  The factor $1/c$ with the speed of light $c\simeq 137$ and the electron charge $q=-1$ are absorbed into the vector potential.}
\begin{eqnarray}
\mathbf{A}(t) = -\int_0^t q \mathbf{E}(t')dt \ .
\end{eqnarray}
Here $\mathbf{E}(t)$ is the electric field vector.
In a typical attosecond streaking or a RABBITT experiment, the target
atom or molecule is exposed to a combination of the two fields:
\begin{eqnarray}
\mathbf{A}(t)= \mathbf{A}_{\rm XUV}(t) + \mathbf{A}_{\rm IR}(t-\tau)
\ ,
\end{eqnarray}
where $\tau$ is the relative displacement of the XUV and IR pulses.
We model an ultrashort XUV pulse by a Gaussian envelope
\begin{eqnarray}
\mathbf{A}_{\rm XUV}(t)=-\mathbf{n}_{\rm XUV}A_{\rm XUV}\exp\left(-2\ln 2
\frac{t^2}{\tau_{\rm XUV}^2}\right)\cos\omega_{\rm XUV} t \ ,
\end{eqnarray}
with the  FWHM $\tau_{\rm XUV}$. The IR pulse is described by the
$\cos^2$ envelope
\begin{eqnarray}
\mathbf{A}_{\rm IR}(t)=-\mathbf{n}_{\rm IR}A_{\rm IR} \cos^2(\pi t/\tau_{\rm IR})
\cos\omega t,\, |t|<\tau_{\rm IR}/2 \ ,
\end{eqnarray}
where $\tau_{\rm IR}$ is the IR pulse duration. The time evolution of
the target under consideration starts from the initial state
\begin{eqnarray}
\Psi(\vecr,t_0)=\varphi_0(\vecr)\exp(-iE_0t_0) \ , 
\label{Psi_init}
\end{eqnarray}
where $t_0=-\tau_{\rm IR}/2+\tau$ and  $\varphi_0(\vecr)$,  $E_0$ 
are the wave function and the energy of the initial state.

After the end of the XUV pulse, the ionized electron is exposed to a
slow varying IR field and the long range Coulomb field of the residual
ion. During the propagation in the IR field,
the photoelectron gains a considerable speed and travels large
distances from the parent ion. To describe this process, solution of
the TDSE should be sought in a very large coordinate box for a very
long propagation time which places a significant strain on
computational resources. To bypass this problem, we employ an
expanding coordinate system \cite{PhysRevA.75.012715}. In this method,
which we term the time-dependent scaling (TDS), the following variable
transformation is made:
\begin{eqnarray}
\vecr = a(t)\bxi \ . 
\label{r_xi}
\end{eqnarray}
Here  $a(t)$ is a scaling factor with an asymptotically linear time
dependence $a(t\to\infty)=\dot{a}_\infty t$ and $\bxi$ is a
coordinate vector. Such a transformation makes the coordinate frame to
expand along with the wave packet. In addition, the following
transformation is applied to the wave function
\begin{eqnarray}
\Psi(a(t)\bxi,t)=\frac{1}{[a(t)]^{3/2}}
\exp\left(\frac{i}{2}a(t)\dot{a}(t)\xi^2\right)\psi(\bxi,t). \label{Psi_psi}
\end{eqnarray}
Such a transformation removes a rapidly oscillating phase factor from
the wave function in the asymptotic region \cite{PhysRevA.75.012715}.  Thus
transformed wave function satisfies the equation
\begin{eqnarray}
i\frac{\partial\psi(\bxi,t)}{\partial
t}&=&\left[\frac{\hat{p}_\xi^{\,2}}{2[a(t)]^2}-\frac{\mathbf{A}(t)\cdot\hat{\mathbf{p}}_\xi}{a(t)}+U[a(t)\bxi]
\nonumber\right.\\ &&\hspace*{-0.5cm}\left.
+\frac{a(t)\ddot{a}(t)}{2}\xi^2-\frac{\dot{a}(t)}{a(t)}\mathbf{A}(t)\cdot\bxi\right]\psi(\bxi,t)
,
\label{TDSExi}
\end{eqnarray}
where 
$\hat{\mathbf{p}}_\xi=-i\nabla_\xi=
-i\left(\frac{\partial}{\partial\xi_x},\frac{\partial}
{\partial\xi_y},\frac{\partial}{\partial\xi_z}\right)$.
Remarkable property of the expanding coordinate system is that the
ionization amplitude $f(\mathbf{k})$ is related with the wave function
$\psi(\bxi,t)$ by a simple formula \cite{PhysRevA.75.012715}
\begin{eqnarray}
|f(\mathbf{k})|^2 = \dot{a}_\infty^{-3} \lim_{t\to\infty}
|\psi(\mathbf{k}/\dot{a}_\infty,t)|^2.
\end{eqnarray}
In practice, the evolution is traced for a very large time $t_f\gg
\tau_{\rm IR}$ and then the ionization probability density is obtained
from the expression 
\be P^{(3)}\equiv\frac{dP}{dk_xdk_ydk_z} = |f(\mathbf{k})|^2 \simeq
\dot{a}_\infty^{-3} |\psi(\mathbf{k}/\dot{a}_\infty,t_f)|^2.  \ee

The coordinate frame \eref{r_xi} is well suited for approximating an
expanding wave packet. However, its drawback is that the bound states
are described progressively less accurately as the coordinate frame
and its numerical grid expands. 
Therefore, during the XUV pulse, when an accurate approximation
of the bound states is required, we use a stationary coordinate
frame. The expansion of the frame starts at the moment $t_1\gg
\tau_{\rm XUV}$.
We use the piecewise linear scaling 
\begin{eqnarray}
a(t)=\left\{
\begin{array}{ll}
1,& t<t_1;\\
\dot{a}_\infty t,& t>t_1.
\end{array}
\right. \label{at}
\end{eqnarray}
At $t<t_1$ the wave function $\psi(\bxi,t)=\Psi(\vecr,t)$.  Since
the time derivative of $a(t)$ defined by \Eref{at} have discontinuity at the
start of the expansion, the wave function at $t_1$ should be
multiplied by the phase factor
\begin{eqnarray}
\psi(\bxi,t_1+0)=\exp\left(\frac{i}{2}\dot{a}_\infty\xi^2\right)\psi(\bxi,t_1-0).
\end{eqnarray}
Here we choose $\dot{a}_\infty=1/t_1$. Such a choice ensures that the
wave packet remains stationary in the expanding frame at $t>t_1$.

Bound states are suppressed by introducing an imaginary
absorbing potential near the origin:
\begin{eqnarray}
U_{sa}(\xi,t)=-i\frac{s(t)}{a(t)}e^{-\xi^2}
\end{eqnarray}
Unlike in the previous treatment \cite{PhysRevA.93.063417}, here we
use a smooth switching of the  imaginary absorbing potential, by setting
\begin{eqnarray}
s(t)=\left\{
\begin{array}{ll}
0,& t<t_1;\\
(1+\cos[\pi(t-2t_1)/t_1])/2,& t\in [t_1,2t_1];\\
1,& t>2t_1;
\end{array}
\right. \label{st}
\end{eqnarray}
This reduces spurious transitions from the bound states to continuum.

After having obtained the ionization amplitude for each individual XUV
pulse in the APT, we summed the results. 
This procedure is correct, because the intensity of the XUV field in RABBITT experiments is usually small, and the processes of sequential absorption of several XUV photons can be neglected.

\subsection{Density~functional~theory~with self-interaction correction}

We employ the density functional theory (DFT) with the
self-interaction correction (SIC) \cite{PhysRevB.23.5048}.  This
correction is necessary to restore the Coulomb asymptotics of the
photoelectron interaction with the residual ion which is essential for
time delay calculations.

The density functional with the SIC \cite{PhysRevB.23.5048} 
contains the Hartree $E_\text{H}\{\rho\}$ and the
exchange-correlation $E_\text{XC}\{\rho_\uparrow,\rho_\downarrow\}$
components:
\begin{eqnarray}
E_\text{SIC}=E_\text{H}\{\rho\}+E_\text{XC}\{\rho_\uparrow,\rho_\downarrow\}-[E_\text{H}\{\rho_i\}+E_\text{XC}\{\rho_i,0\}]
 , ~~
\end{eqnarray}
where the electron density
\begin{eqnarray}
\rho(\vecr)=\sum_{i=1}^{N_e}\rho_i(\vecr)
\end{eqnarray}
is the sum of  the particle densities of the $i$-th electron orbital
\begin{eqnarray}
\rho_i(\vecr)=|\varphi_i(\vecr)|^2 \ .
\end{eqnarray}
Here we consider only molecules with fully coupled electrons and hence
the density of the spin-up and spin-down electrons are equal:
\begin{eqnarray}
\rho_\uparrow(\vecr)=\rho_\downarrow(\vecr)=\rho(\vecr)/2
\end{eqnarray}
The Hartree energy
\begin{eqnarray}
E_\text{H}\{\rho\}=\frac{1}{2}\int \int \frac{\rho(\vecr)\rho(\vecr')}{|\vecr'-\vecr|} d\vecr' d\vecr
\end{eqnarray}
The exchange-correlation functional is expressed in the local density
approximation (LDA)
\begin{eqnarray}
E_\text{XC}\{\rho_\uparrow,\rho_\downarrow\}=\int
[\rho_\uparrow(\vecr)+\rho_\downarrow(\vecr)]
\varepsilon_\text{XC}[\rho_\uparrow(\vecr),\rho_\downarrow(\vecr)]
d\vecr ,
\end{eqnarray}
where
$\varepsilon_\text{XC}[\rho_\uparrow(\vecr),\rho_\downarrow(\vecr)]$
is the exchange-correlation energy per an electron. The effective
potential acting upon an $i$-th electron by the rest of the
many-electron ensemble is expressed as a functional derivative
\begin{eqnarray}
u_i(\vecr)=\frac{\delta E_\text{SIC}}{\delta\rho_i(\vecr)}
\end{eqnarray}
The Kohn-Sham effective potential is the sum of the nuclear and
electron components:
\begin{eqnarray}
U(\vecr)=u_\text{nucl}(\vecr)+u_i(\vecr)
\end{eqnarray}
The SIC can be applied to any density functional. 
In the present application, we used LDA exchange-correlation functional proposed in \cite{PhysRevB.13.4274}.

Calculation of the one-electron orbitals $\varphi_i(\vecr)$ and
corresponding one-electro potentials $u_i(\vecr)$ is carried out
by the imaginary time evolution based on the solution of the equation
\begin{eqnarray}
\label{DFT}
-\frac{\partial\varphi_i(\vecr,t)}{\partial
t}=\left[\frac{\hat{p}^{\,2}}{2}+u_\text{nucl}(\vecr)+\alpha_k u_i(\vecr)\right]\varphi_i(\vecr,t)
\end{eqnarray}
The orthogonality to the occupied states
is enforced on each step of the time evolution:
\begin{eqnarray}
\varphi_i(\vecr,t+0)=\varphi_i(\vecr,t)-\sum_{j=1}^{i-1}\langle
\varphi_j(\vecr) | \varphi_i(\vecr,t) \rangle \varphi_j(\vecr)
\end{eqnarray}
The stationary orbital is evaluated as the limit
\begin{eqnarray}
\varphi_i(\vecr)=\lim_{t\to\infty}\varphi_i(\vecr,t)
\end{eqnarray}
After finding $\varphi_i(\vecr)$, a new set of potentials
$u_i(\vecr)$ is determined and fed into \Eref{DFT} to start the
next iteration. On the first iteration, the parameter $\alpha_1=0$,
i.e. only the nuclear term is taken into account in \Eref{DFT}. On the
next iterations, $\alpha_{k>1}$ grows linearly with $k$ reaching the
value of $\alpha_{k\gg1}=1$ . This way the inter-electron interaction
is switched on gradually thus ensuring a smooth convergence of the
solution.

The orbital energies are calculated as 
\begin{eqnarray}
\epsilon_i=-\frac{1}{2}\lim_{t\to\infty} \left[ 
\frac{1}{\langle\varphi_i(\vecr,t) | \varphi_i(\vecr,t)\rangle} 
\frac{d\langle
\varphi_i(\vecr,t) | \varphi_i(\vecr,t) \rangle}{dt} \right]
\end{eqnarray}
Thus determined the ionization potential of the \H molecule is equal
to $|\epsilon_1|$=16.7~eV whereas the experimental value is
15.6~eV. The calculated ionization potentials $|\epsilon_i|$ of the
\water molecule are shown in \Tref{tab:H2Oenergies} in comparison with
the experimental values from \cite{NING200819}.

\begin{table}
\centering
\caption{Calculated and experimental ionization potentials of the
  \water molecule}
\begin{tabular}{|c|c|c|}
\hline
Orbital & DFT-SIC & Expt.\cite{NING200819}\\
\hline
$1b_1$ & 12.3 & 12.6 \\
$3a_1$ & 14.9 & 14.8 \\
$1b_2$ & 18.3 & 18.7 \\
$2a_1$ & 34.3 & 27.1 \\	
\hline		
\end{tabular}
\label{tab:H2Oenergies}
\end{table}

To solve the TDSE \eqref{TDSExi} we employed the one-electron
effective potential $u_i(\vecr)$ corresponding to the ground
stationary state and this potential was frozen during the whole
ionization process (the frozen-core approximation).

\subsection{Numerical implementation}

We solve the TDSE.~\eref{TDSExi} using the orthogonal fast spherical
Bessel transform as described in \cite{Serov201763}.
In all the calculations, we set the box size to
$\xi_{max}=51.2$~a.u. The radial grid step was set to $\Delta
\xi=0.2$~a.u. The angular basis parameters were $N_\theta=16$ and
$N_\phi=8$ for H$_2$, and $N_\theta=4$ and $N_\phi=6$ for H$_2$O.
The APT is modeled by a series of $N_{\rm APT}=11$ Gaussian pulses
with the width $\tau_{\rm XUV}=1$ a.u. (24~as). Such a short pulse
duration leads to a large spectral width and allows to obtain the time
delay results in a wide range of photoelectron energies.

The APT width $\tau_{\rm APT}=2T_{\rm IR}$ (5.2~fs), whereas a long IR
pulse is modeled by a continuous wave with the frequency
$\omega=0.05841$ a.u. (photon energy 1.59~eV, $\lambda=780$~nm) and
the vector potential amplitude $A_{\rm IR}=0.025$.  The amplitude of
the XUV pulse was $A_{\rm XUV}=0.2$~a.u.  The relative APT/IR time
delay $\tau$ was varied from 0 to 0.5$T_{\rm IR}$ with a step
0.03125$T_{\rm IR}$.
The central frequency was $\omega_{\rm XUV}=29\omega$ for H$_2$ and $\omega_{\rm XUV}=39\omega$ for water.

By exposing an atom or a molecule to the APT with the
central frequency $\omega_{\rm XUV}=(2q_{0}+1)\omega$, the photoelectrons
will be emitted with the energies
$E_{2q+1}=(2q+1)\omega-E_0$ 
corresponding to the odd harmonics of the IR frequency $\omega$. Superimposing a dressing IR field will add additional peaks in the photoelectron spectrum at 
$E_{2q}=2q\omega-E_0$
These additional peaks, known as the sidebands (SB), correspond to the
even harmonics. The sideband amplitudes will vary with the relative
time delay $\tau$ of the APT and the IR pulses as \cite{Paul01062001}
\begin{eqnarray}
S_{2q}(\tau)=A+B\cos[2\omega(\tau-\tau_a)], 
\label{SBtau}
\end{eqnarray}
where $\tau_a$ is the atomic time delay.  
The atomic time delay can be written in a form
\be
\tau_a = \tau_{\rm W} + \tau_{\rm CLC}
\ ,
\label{atomic}
\ee
where $\tau_{\rm W}$ is the Wigner time delay \cite{PhysRev.98.145}
and $\tau_{\rm CLC}$ is the Coulomb-laser coupling (CLC) correction
\cite{0953-4075-44-8-081001}.
Here we assume that there is no group delay (chirp) in the APT spectrum and all the
harmonics have the same phase.

\begin{figure}[ht]
\includegraphics[angle=-90,width=0.8\columnwidth]{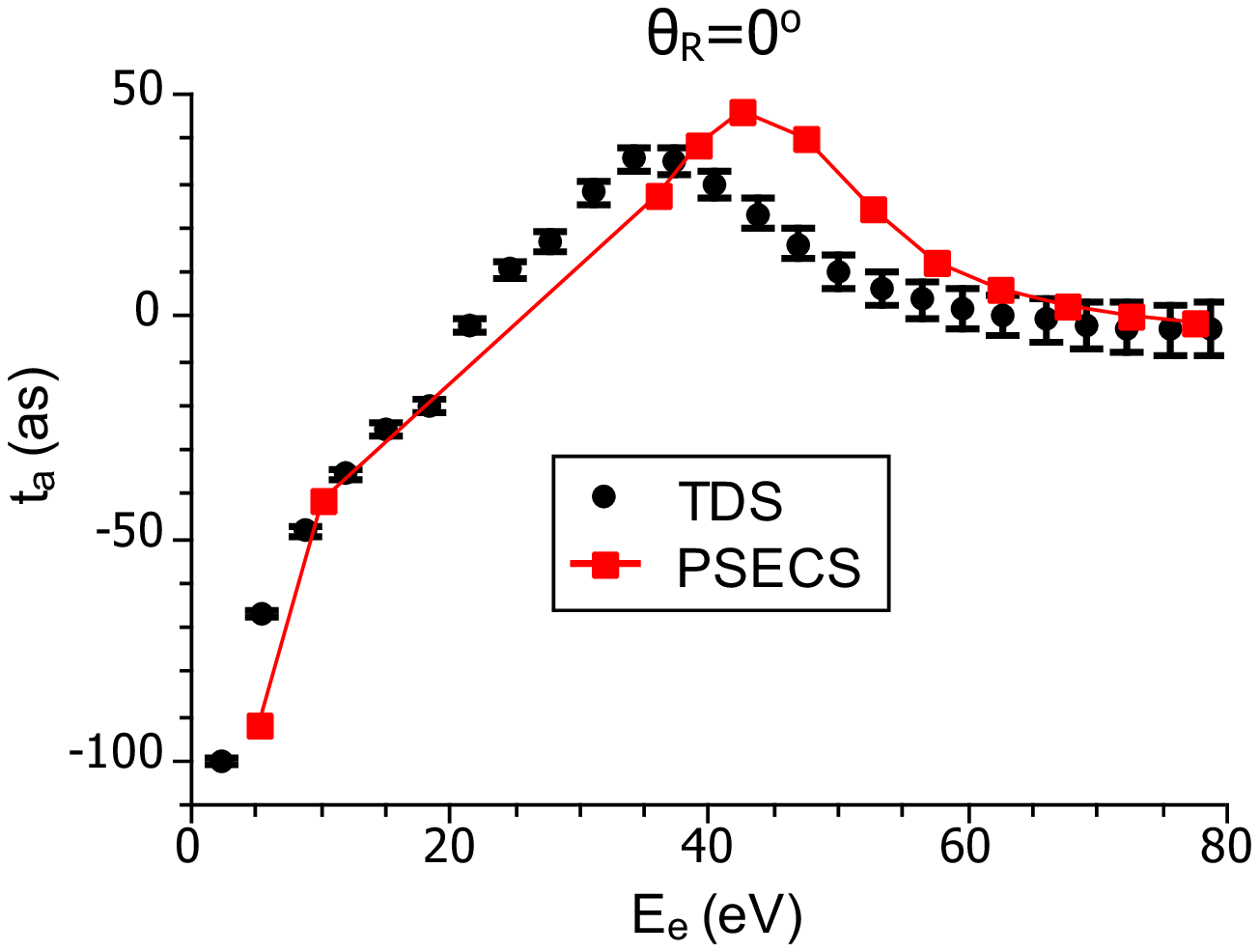}
\includegraphics[angle=-90,width=0.8\columnwidth]{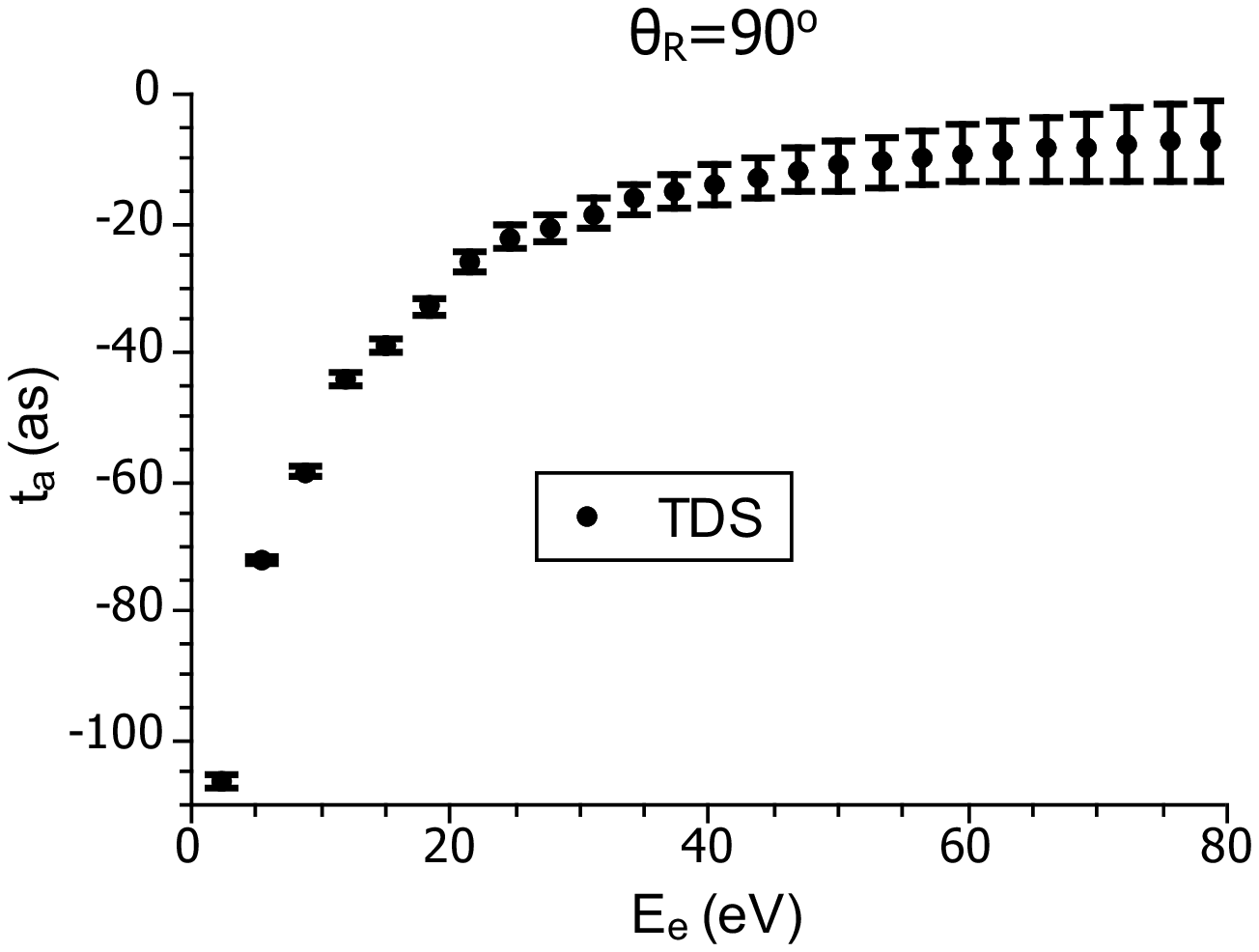}

\caption{Atomic time delay of \H as a function of the photoelectron
  energy $E_e$ for emission in the polarization direction.  The
  molecular axis is aligned along (top panel) and perpendicular
  (bottom panel) to the polarization direction}
\label{fig:H2}
\end{figure}

\section{Results}

In \Fref{fig:H2} we display the atomic time delay calculated by the
TDCS method. Every energy point on the graph corresponds to a given
SB. The error bars indicate the accuracy of the cosine fit to 
\Eref{SBtau}. For comparison, another calculation is shown in which
the Wigner time delay is calculated by the prolate spheroidal exterior
complex scaling (PSECS) \cite{PhysRevA.87.063414} and the CLC
correction is introduced analytically \cite{Serov2015b}. The PSECS is
an {\em ab initio} technique and it returns the exact Wigner time delay
for diatomic molecules. However, in the photoelectron energy range 
between  10 and  36~eV, the PSECS calculation on \H did not
converge. Most likely, this is because of a large number of
quasi-stationary states in this spectral range.

The TDCS and PSECS+CCLC results agree very well close to the threshold
and are qualitatively similar at large excess energies. In the
parallel molecular orientation, both set of calculations display a
peak in the atomic time delay. However, in the TDCS calculation this
peak is shifted by 7~eV towards lower photoelectron energies
($E_e=35$~eV in TDCS versus 42~eV in PSECS+CCLC). Such a large
difference can be explained by poor performance of the DFT for such a
few-electron systems like \H. We note that the peak displacement by
7~eV far exceeds an error of 1~eV in the ionization potential. This
indicates that such a dynamic quantity as the atomic time delay is
much more sensitive to inter-electron correlation than the static
ionization potential.

\begin{figure}[ht]
\includegraphics[angle=-90,width=0.8\columnwidth]{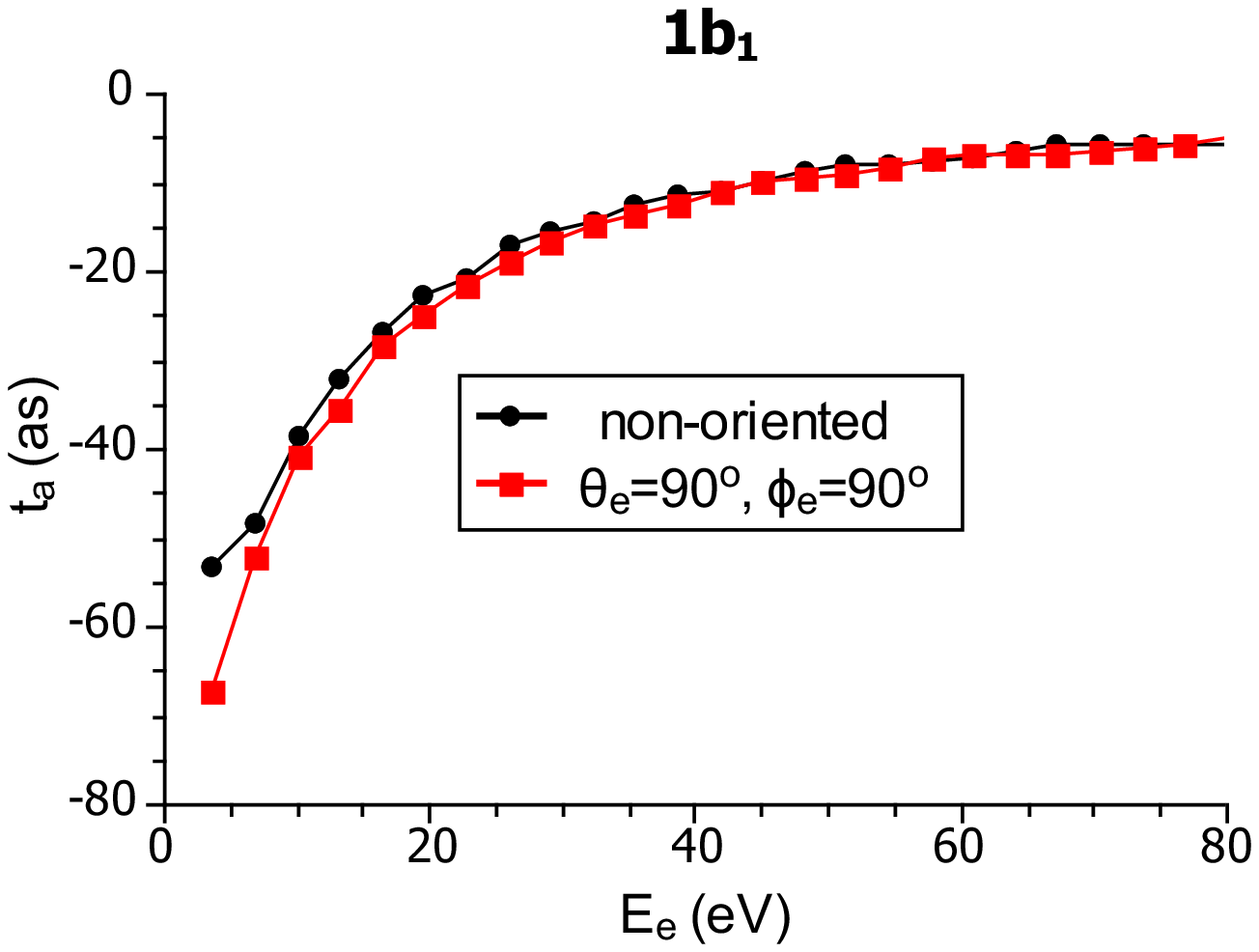}
\includegraphics[angle=-90,width=0.8\columnwidth]{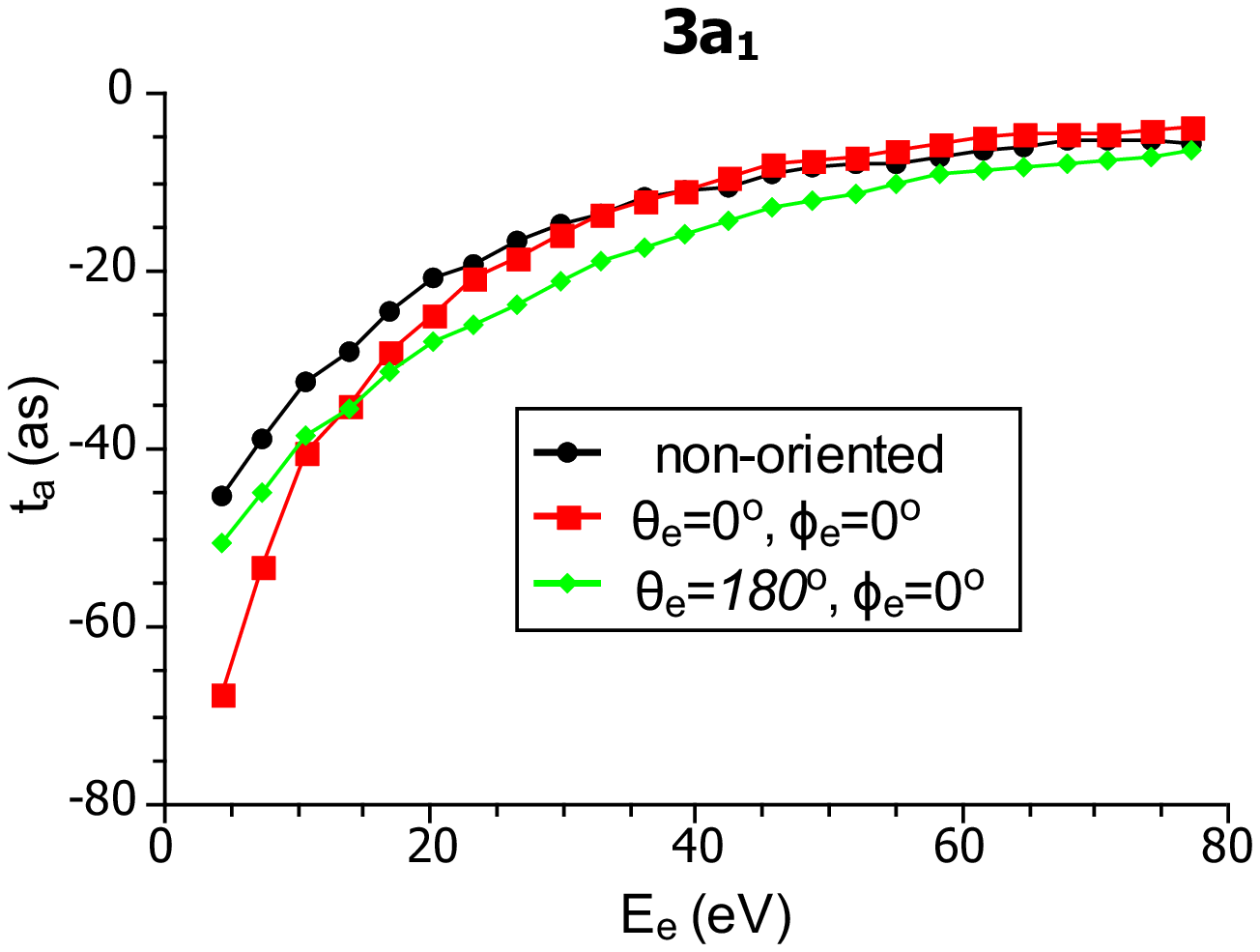}

\caption{Atomic time delay in \water  for randomly oriented
  and aligned molecules. The top panel - the 
$1b_1$ orbital, the bottom panel the  $3a_1$ orbital.}
\label{fig:H2O}
\end{figure}

\Fref{fig:H2O} shows the atomic time delay in \water molecule
corresponding to ionization of the HOMO $1b_1$ and the HOMO-1
$3a_1$. Various plotting symbols display the time delay for randomly
oriented and aligned molecules. The randomly oriented results are
averaged over all the molecular orientations. The aligned results
correspond to photoelectron emission in the direction perpendicular to
the nodal plane of the given orbital. In the case of $1b_1$ this plane
contains the H atoms and the molecular dipole momentum vector. Therefore both
perpendicular orientations produce identical results. The nodal plane
of $3a_1$ is perpendicular to the dipole momentum and hence there are
two distinctive perpendicular orientations: in the direction of the H
atoms ($\theta=0^\circ$) and the reverse direction ($\theta=180^\circ$).

In the case of the $1b_1$ orbital, the randomly oriented time delay is
nearly identical to the aligned results as the perpendicular emission
is by far dominant.
In the case of the $3a_1$ orbital, results are more complicated.  At
large photoelectron energies, the atomic time delay for $\theta=180^\circ$
is large than that for $\theta=0^\circ$ which means the escape is faster in
the O direction than in the H direction. However, this trend is
reversed near threshold. Away from the threshold, the non-aligned time
delay is in between these two cases whereas it is above both the
aligned curves close to the threshold. This means that some other
orientations contribute strongly.

\begin{figure}[ht]

\includegraphics[angle=-90,width=\columnwidth]{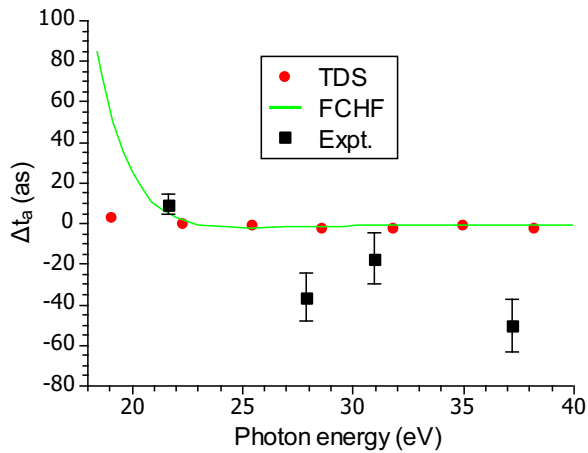}

\caption{Time delay difference between the $3a_1$ and $1b_1$ orbitals
in the \water molecule as a function of the photon energy.   The
filled (red) circles -- TDCS calculation, the (green) solid line - the
FCHF calculation \cite{PhysRevLett.117.093001,Baykusheva2017}, the filled squares with error bars - experiment
\cite{PhysRevLett.117.093001} }
\label{fig:H2O_Expt}
\end{figure}

In \Fref{fig:H2O_Expt} we display the time delay difference between
the $3a_1$ and $1b_1$ orbitals at the same photon energy. It may seem
surprising that the time delay difference is nearly vanishing in the
whole studied energy range.  This, however, may be explained by the
fact that the randomly oriented water molecule may look like the neon
atom and these two states in Ne differ only by the nodal plane
orientation. 
In the same figure we plot the experimental results by
\citet{PhysRevLett.117.093001} which show a sign variation of the time
delay difference. The frozen-core Hartree--Fock (FCHF) calculation by
the same authors \cite{PhysRevLett.117.093001,Baykusheva2017} is also
overplotted. Neither calculation reproduce the experimental results
within the stated error bars.

\section{Conclusions}

We combined the time-dependent coordinate scaling (TDCS) method, which
we developed earlier for modeling of RABBITT experiments 
\cite{PhysRevA.93.063417}, with the density functional
theory with self-interaction correction (DFT-SIC). An advantage of the
TDCS method is the direct solution of the time-dependent Schr\"odinger
equation driven by the superposition of the XUV and IR pulses without
the splitting of the atomic time delay into the Wigner and CLC
components.  By using this technique, we calculated the
photoionization time delay of the \H and \water molecules. In the case
of \H, we made a comparison with the {\em ab initio} PSECS
calculations and found a good agreement. In the case of \water, a
comparison was made with experiment and agreement was found poor, but
other theoretical methods like FCHF did not perform any better.

The theory consistently point to nearly identical time delay from the
HOMO and HOMO-1 orbitals of the randomly oriented water
molecule. However, this time delay difference becomes noticeable on
the oriented \water molecule. Hence, the experiments will be highly
desirable in this oriented configuration which may reveal a reach and
anisotropic ultrafast photoelectron dynamics.



\vfill\eject \np


\end{document}